# A SUPERCOMPUTING BASED DISTRIBUTED CLOUD MARKETPLACE.

## v.2.5.0

**LuluChain Team**: Minjun Kim and Sina Falaki

*September 14, 2018*

**Abstract:** The once mythological 51% attack has moved beyond the hypothetical and now poses a legitimate widespread threat to blockchain technology. Current blockchains provide inferior throughput capacity when compared to that of centralized systems, creating an obvious vulnerability which allows the 51% attack to occur within decentralized systems. Despite recent advancements in blockchain which introduce interesting models that achieve high throughputs with enhanced security and privacy, no current networks have evolved to deploy the optimal solution of combining scalability, security, and distributed systems to create a legitimate supercomputing enterprise-grade developer sandbox. In this paper, we introduce an infinitely scalable, secure, and high throughput blockchain capable of amassing supercomputer speeds with off-the-shelf hardware, LuluChain. LuluChain simplifies the blockchain model to obtain greater functionality, speed, scalability, privacy, and flexibility, that works to combat the inflated pricing models set by the oligopolistic cloud computing market as it requires minimal computational work. By eliminating the need for timestamp synchronization and majority agreement among all participants, LuluChain opens the door to reliable trust, low-cost instant transactions, and flexible instant smart contracts. The supercomputing, high throughput distributed system is the ideal foundation for an essential distributed cloud marketplace.

## 1. INTRODUCTION

LuluChain is a new algorithmic plug and play blockchain architecture with a decentralized supercomputing layer. It functions as a highly optimized chain based off a distributed economy model allowing for high throughput and scalability via ledger separation with todays off-the-shelf hardware. To accelerate consensus-based decision making, LuluChain maintains a separate ledger for each user - rather than amassing all transactions into a single, often clunky global ledger. This extremely fast transaction network makes for realistic present-day integration with the help of a Distributed Bridging Propagation System. This serves as the framework for a fee-less distributed economy where both user and governing body receive equal payment for their work in Lulu Cash. With an added layer of quality control, streamlined security and certificate protocols in place LuluChain acts as a developer oasis. With our plug and play smart contract SDK and API's, an abundant of use cases are possible with LuluChain. Decentralized Ex-changes, Ad Networks, and the development and integration of Dapps. The platform serves to eliminates developer worries of gathering enough computing resources, thus affording them the freedom to focus on building innovative and high-quality blockchain applications with virtual ease.

**Note:** LuluChain is in development. Our research and new versions of this paper will appear at https://LuluChain.org.

## 2. BACKGROUND

Satoshi Nakamoto published his whitepaper Bitcoin: A Peer-to-Peer Electronic Cash System in 2008, outlining the basis for what is known today as blockchain and inspired a technological revolution. Since Nakamoto solved the problem of the double-spend with the public ledger system, blockchain has flourished as a development platform for various applications. From financial record keeping to supply chain management, the adoption of blockchain as a fundamental technology has acted like a drop of ink in a glass of water, permeating all industry landscapes. As developers and businesses have begun to integrate blockchain, they have begun to experience the varying issues that plague decentralized tech: lack of scalability in addition to privacy and security problems.With Nakamoto consensus, all nodes agree upon the ordering of transactions by accepting a "valid" block (found via either proof-of-work or proof-of-stake) and waiting for it to propagate through the network and obtain a sufficient number of confirmations. Once enough confirmations have been achieved, the game theoretic argument is that honest participants will continue to secure the longest chain. As long as honest participants remain in the majority, it becomes increasingly difficult for a block with many confirmations to be excluded from the longest chain in the event of a fork. However, in practice, this scheme still poses issues within a non-synchronized network. In a non-synchronized network, several consensus participants can create a block at nearly the same time due to latency and network topology. Subsets of the network receive and propagate their version of the "valid" block to peers which then leads to a fork. A fork will continue until the network agrees on which branch to extend according to the longest-chain rule. When this occurs, other branches will then be pruned and discarded. Forks cause network inefficiencies. Every time a fork occurs and a "validated" block gets discarded by the forked portion of the network that has approved the block. This results in the system slowing down unnecessarily because each block must fully propagate before a new one is created, therefore unless a fork occurs and the forked network discards some subchain in accordance to the longest chain rule. Blocks must be generated regardless of whether any updates in the database occur. This creates a dilemma that not only requires storage space, but such an event also wastes computing resources such as bandwidth and processing power. In contrast to this PoW (Proof of Work) system, where the consensus power is proportional to the amount of computing power, PoS (Proof of Stake) depends on the amount of funds owned by a single user. Running PoS consensus protocols requires little computational power and makes it relatively simple to implement a 'one coin - one vote' rule to prevent centralization. 51% attacks and network overloads which were once reserved for hush-toned catastrophizing sessions are now discussed widely in the blockchain community on a regular basis. Bitcoins' network inefficiencies have become focal conversation across every channel. Massive increases in issued transactions and growing user adoption has ultimately led to the demise of using Bitcoin as a practical payment solution. 51% attacks on Monacoin, Bitcoin Gold, ZenCash, Verge, and ultimately; Litecoin Cash, have shaken the blockchain world. A technology initially viewed as a decentralized savior now requires much innovation and saving itself.

## 3. LEDGER AND NETWORK

Cryptographic distributed ledgers are becoming widely used for monetary transactions, and are poised for acceptance in execution of smart contracts. Bitcoin and Ethereum are the leaders among thousands of ventures in the field. However, all current cryptographic distributed ledger technologies are held back from wider adoption by several practical problems:

· **Speed.** Current algorithms for establishing agreement on the content of the ledger cause the computers to consume significant time and electricity to prove and communicate agreement. These costs are passed to users in the form of transaction fees and mintage inflation. Current algorithms do not directly scale to the transactions rate that would compete with existing, centralized banking and credit card systems.

· **Complexity.** A cryptographic ledger may be distributed across thousands or millions of computers around the world, with no central authority. Because any of the computers may be compromised and controlled by an adversary, complex consensus algorithms are used to ensure that compromised computers cannot steal and cannot affect the ledger.

· **Privacy.** Most current cryptocurrencies use account pseudonyms, in validating the correctness and integrity of each transaction. Although an account pseudonym itself does not reveal the user's name, transactions can be linked to one another via pseudonyms, IP addresses, and timestamps, narrowing the set of possible users; individual transactors can sometimes be identified. Thus privacy in most cryptocurrency systems is weak. As a result, recipients of large transactions may fear targeting by criminals. Businesses may fear exposing their transaction history to competitors.

· **Laundering.** Some users may exploit the pseudonymity of distributed ledger systems to create a series of transactions that obfuscates the origin of their funds, or that moves funds abroad. This behavior may contravene laws in some jurisdictions, and may complicate efforts to track criminal activities.

Solving these issues is the aim of intensive research and development efforts, toward full realization of the potential benefits of cryptographic distributed ledger systems, which include:

· **Secure.** The ledger protects against fraud and theft. There are no paper certificates or notes to manage.

Transactions are controlled to ensure that no account becomes overdrawn, that deposits equal receipts, and that no funds are misrouted.

· **Decentralized.** The operation and governance of the ledger are invulnerable to a single point of failure.

· **Permanent.** The ledger cannot be terminated or destroyed by any party. For example, participants need not worry about a central provider dropping support for a no-longer profitable ledger service.

· **Neutral.** The ledger structure and operation are aligned to serve its participants, not a service provider. Users can trust that they have equal status to one another, independent of profit, politics, and power.

· **Open.** The system's underlying code, structure, and data are inspectable and verifiable at any time. Operations are transparent and hence trustable.

· **Universal.** The ledger operates across borders, without requiring different rules in different locales.

· **Liberating.** The ledger allows actions that circumvent restrictions of repressive, authoritarian, or intolerant regimes.

· **Uncorrupt.** The ledger architecture provides convincing algorithmic guarantees of system integrity, and inspectable logs of activity.

· **Private.** The ledger can protect participants from some forms of spying, wiretapping, and disclosure.

· **Apolitical.** Authorities can't intrude or interfere on the basis of governmental will. Participants are protected from repression and censorship, and assets are protected from seizure.

· **Objective.** Distributed ledger systems and smart contracts are based on well-defined data and measurements, with no opportunity for subjective interpretation.

· **Available.** Around the clock and around the globe.

· **Inexpensive.** Distributed ledgers remove transaction middlemen, eliminate clearing and settlement infrastructure, eliminate interbank and international interfaces and fees, and reduce errors and exceptions. Thus they can reduce transaction overhead costs.

### LuluChain, the parallel chains

The original Bitcoin paper by Satoshi Nakamoto [1] proposed a data representation for transactions that has been used in most subsequent distributed ledger technologies. The data representation comprises an "electronic coin" that can be transferred from owner to owner by adding digital signatures in a chain. An important consequence of choosing this data representation in a distributed ledger is that transactions must be validated via a distributed agreement procedure, across all computer nodes in the distributed network. The agreement procedure ensures consistency of the ledger, even when some of the nodes may have been compromised by an attacker. The agreement establishes an ordering of transactions. Given an agreed ordering, an owner's attempt to spend a given coin a second time can be detected as fraudulent and rejected - thus ensuring integrity of the ledger.

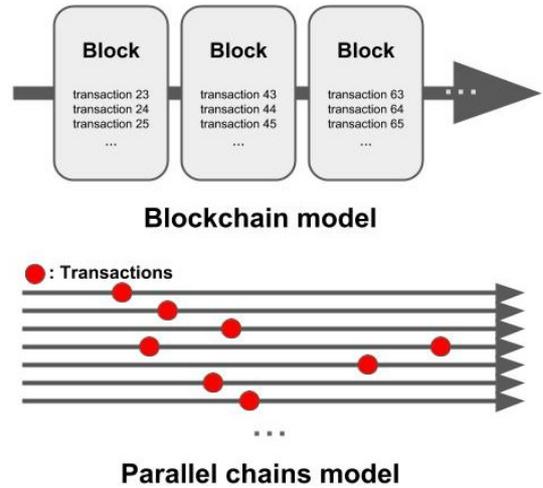

**Fig. 1.** Chain model comparison between blockchain and parallel chains.

A key feature of LuluChain is that transaction ordering information is encoded within each transaction record (See Fig. 1), rather than being subject to post-transaction agreement across all computer nodes as in Bitcoin. This data representation feature allows new benefits: (1) ordering of transactions need not be established globally across all user accounts (as in Bitcoin), but separately for each individual account; and (2) no reversion of previously agreed transactions (as in Bitcoin) is needed. These benefits fundamentally simplify the transaction agreement procedure relative to other distributed ledger systems. As a result, LuluChain is simpler, faster, scales more readily, imposes less computational load, and is more stable than other distributed ledgers. Because LuluChain's transaction data representation allows ordering to be established separately for each account, and recorded in a separate sequential chain for each account, the system is termed parallel chains. This is in contrast to other distributed ledger systems, in which great computational effort must be expended to agree on a total ordering of all transactions across all accounts, into a single chain. Lulu's parallel chains model has the following characteristics:

· Computationally efficient. True consensus greatly simplifies transaction agreement.
· Resilient to 99%-majority attacks. As long as at least one honest node is present, fraud is prevented.
· Skew-free. Eliminates the need to rely on timestamps or synchronized clocks.
· Depends on asymmetric PKI. Security rests in each user's private keys.
· Sybil attack resistant. Extra identities do not help an attacker.
· Partition attack resistant. Cutting off parts of the network does not let separate chains form.

The Hashgraph methods use a representation of multiple chains, but differ in several key ways from the Lulu parallel chains model: In LuluChain, no computations to establish an agreed ordering of transactions or events are performed;

no agreement on timestamps is required; no node can influence another's validation checks for a transaction; no record of how the nodes have communicated is required; no distributed elections, votes, supermajorities, or rounds are required; no forking of ledgers is possible. Because of the simplicity and efficiency of its model, Lulu requires no sharding, delegation, or other workarounds to enhance scalability. Nevertheless, those structures can optionally be implemented within a Lulu network, if needed for other reasons, such as hierarchical segregation of transactions or applications.

***LuluChain True Consensus procedure.*** The Bitcoin distributed agreement procedure is majoritarian, in the sense that an adversary who gains control of 51% of the computing power in the network can then corruptly write to the ledger at will. In contrast, Lulu implements a distributed agreement procedure termed true consensus. To reach true consensus on a proposition, two conditions must be met: a minimum number of participants must support the proposition, and none may oppose. In a LuluChain distributed ledger, true consensus on a proposed transaction is established when a minimum number of computer nodes provide valid evidence for validity of the transaction, and none provide valid evidence for its invalidity. LuluChain includes information in each transaction record which allows any one computer node to invalidate a proposed transaction, by showing valid evidence from a ledger. For example, valid evidence for the invalidity of a double-spend attempt might comprise a duly signed transaction record of a prior spend, alongside the proposed transaction record for a second spend.

***LuluChain protocol.*** In LuluChain, every account is identified by a hash string derived from the public key of the account. Every transaction record is signed by the sender and receiver accounts, using both their private keys. Consequently, any node can verify that both accounts have agreed to the transaction. In LuluChain network, nodes can serve as witness nodes to perform distributed validation services in exchange for incentive compensation. If a node is found to falsify a witness report, the LuluChain network bans the node from further witness eligibility.

A proposed transaction record is broadcast to a set of witness nodes. Each witness node verifies the signatures on the record and checks its ledger for the transaction sender. If a witness node detects an improper signature, a double-spend attempt, or insufficient account balance, it creates a rejection record. Otherwise it creates an acceptance record. The rejection or acceptance record contains the proposed transaction and the relevant evidence: a conflicting transaction record, or a supporting ledger entry. The witness node signs the rejection or acceptance and broadcasts it to the network, and to the transaction sender and receiver nodes. After a specified brief waiting period, a node receiving at least a minimum number acceptance messages for a transaction, and no rejection messages, may safely enter the transaction into its ledgers for the sender and receiver. A node receiving a valid rejection record for a proposed transaction penalizes the transaction sender account. The penalties include refusing to accept some or all future transactions from that sender account (thus "slashing" the value of the account), and broadcasting a rejection for future transaction attempts on that account. A witness node that attempts to falsify a rejection or acceptance, by broadcasting invalid evidence, is also penalized. The penalties include stake forfeit and exclusion from future witness eligibility, as well as rejection of the witness message itself. A node verifies the signatures and evidence for all proposed transaction records and witness acceptances/rejection records, before considering the records valid. Thus false records are not accepted and do not affect the ledger (See Fig. 2). In other distributed ledger technologies, such as Bitcoin and Ethereum, every node must agree on a single, total ordering of all transactions before a transaction can be accepted. Consequently, those technologies are inherently subject to delays that degrade their performance. In contrast, LuluChain nodes achieve correct, secure operation by maintaining a separate ledger for each participant's transactions, without having to order transactions of different participants. Lacking the total-ordering agreement bottleneck, LuluChain is free to run at full network speed.

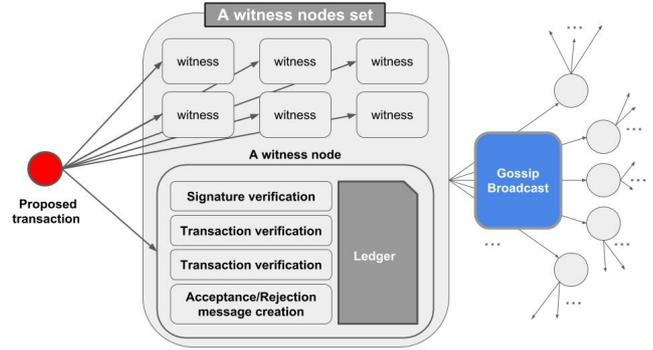

Fig. 2. LuluChain protocol procedure of a proposed transaction. A proposed transaction is sent to a selected set of witness nodes for verifications, and once verified, propagated to all the other nodes through a gossip protocol.

***Witness nodes selection.*** The selection of a set of witness nodes must be performed in a specific way that possible attackers has no chance to predict the set. LuluChain implements the concept of "Random Network" to dynamically construct a network, to configure the global network topology among witness nodes at random, and applies the highly scalable asynchronous fluid community detection algorithm to get a set of detected nodes in a same community which will be the final witness nodes for a specific transaction request. We will discuss the asynchronous fluid community detection algorithm in a later subsection.

A random network is made out of N nodes with edges connecting each pair of nodes with a chosen probability p. Witness node selection module, first, chooses the connection probability p, and constructs a random network with all available N witness nodes. For each and every possible node pair, it generates a random number 0 and 1. If the generated number is greater than the connection probability, make an edge between the two node of the pair, otherwise, there is no edge between them. The module repeats the simple random connection for each of the $N(N-1)/2$ node pair.

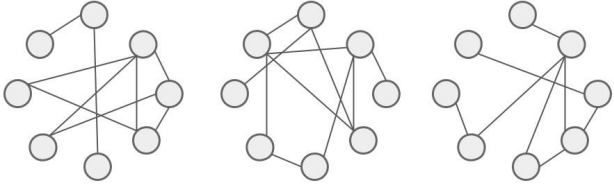

Fig. 3. Consecutively generated three random networks with the same number of nodes and the connection probability. The three random networks show totally different configurations. Hence the topological information of the networks are all different as well.

```
# simplified code representation
# random network
# running on a central node

import random

N = len(all_witness_nodes_list)
# connection probability.
p = random.random()
""" build a list of all node pairs
pairs = [(n1, n2), (n1, n3), ...] """
pairs = get_pairs(all_witness_node_list)
# generate a network with N nodes.
G = network(N)

for pair in pairs:
    prob_threshold = random.random()
    if p > prob_threshold:
        """ make an edge between the two
        nodes of the pair """
        G.connect(pair)
    else:
        if isConnected(pair) == True:
            G.disconnect(pair)
        else:
            pass
```

Because this specific process, even with the same *N* and *p*, generates slightly different network every time when this process is executed, for each proposed transaction, we can expect that any attackers cannot assume/predict which nodes will be the final set of witness nodes for verifying the transaction (See Fig. 3).

Under a decentralized system, because there cannot exist a trustable central node that constructs a random network and propagates the topological information of the network to other nodes, this specific process needs to be localized which means that each node should be able to create their own choice of connections to other nodes, and then forms a global topology that is the same as others. In this case, first, each node($i$) randomly chooses random number($k_i$ from 0, $N(N-1)/2$) of nodes it will be connected to, and get other nodes' selections to build the network adjacency matrix and to construct the globally shared random network.

```
# simplified code representation
# localized random network
```

```
# running on each node

import random

N = len(all_witness_nodes_list)
all = all_witness_nodes_list
# number of nodes to be chosen.
n = random.randint(0,N)
# select n nodes from the all nodes list.
selection = random.sample(all, n)

""" request other nodes' selection
to create the network adjacency list """
selections = []
for node in all:
    r = request(node, selection).read()
    selections.append(r)
# make the adjacency list with selections.
adjacency_list = make_adj_list(selections)

""" construct the random network
with the adjacency list """
G = creat_network(adjacency_list)

""" detect communities with
asynchronous fluid algorithm
and select the largest community """

communities = asynchronous_fluid(G)
selected_community = max(communities)

# define the final set of witness nodes.
witnesses = selected_community.nodes()

""" decide whether or not to
take the proposed transaction's
verifications steps """

# get my identification.
I = config.id()
if I in witnesses:
    proceed to transaction verifications
else:
    do nothing
```

Once the specific random network is formed, every witness node of the network has the global network topology on which the asynchronous fluid community detection algorithm is applied. Among all the communities, LuluChain uses the largest community and its components nodes as the final witness nodes to serve the transaction verification works.

***Asynchronous Fluid Communities.*** To detect possible communities of our random network, LuluChain implements the Asynchronous Fluid Communities Algorithm. Fluid Communities (See. Fig. 5) is a network community detection algorithm which is based on the propagation methodology and low in computational cost and high scalability. While being highly efficient, Fluid Communities can cluster network nodes in graphs based on properties of the nodes, and also are capable of identifying a variable number of communities within the network.

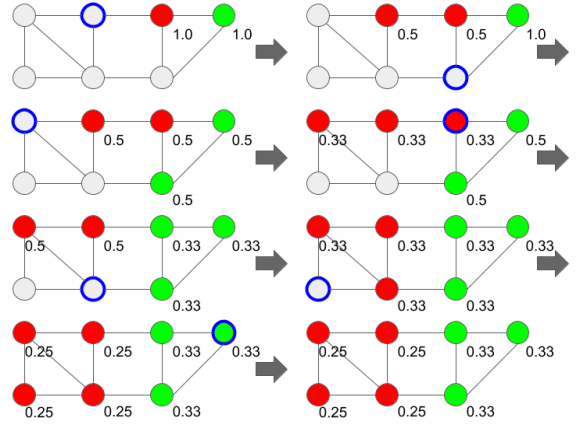

Fig. 5. Workflow of Fluid Communities for k=2 communities (red and green). Each node assigned to a community is labeled with the density of that community. The update rule is evaluated on each step for the vertex highlighted in blue. The algorithm converges after one complete superstep.

$$\delta(c(w), c) = \begin{cases} 1, & \text{if } c(w) = c, \\ 0, & \text{if } c(w) \neq c. \end{cases} \tag{3}$$

where $v$ is the mode being updated, $\varepsilon'_v$ is the set of candidates to be the new community of $v$, $\Gamma(v)$ are the neighbours of $v$, $d(c)$ is the density of community $c$, $c(w)$ is the community node $w$ belongs to and $\delta(c(w), c)$ is the Kronecker delta. Notice that $\varepsilon'_v$ contains the current community of the node $v$, $v$ does not change its community. however, if $\varepsilon'_v$ does not contain the current community of $v$, the update rule chooses a random community within $\varepsilon'_v$ as the new community of $v$. This completes the formalization of the update rule (4). :

$$c'(v) = \begin{cases} x \sim v(\varepsilon'_v), & \text{if } c(v) \notin \varepsilon'_v, \\ c(v), & \text{if } c(v) \in \varepsilon'_v. \end{cases} \tag{4}$$

where $c'(v)$ is the community of node $v$ at the next superstep, if $\varepsilon'_v$ is the set of candidate communities from equation 2 and $x \sim v(\varepsilon'_v)$ is the random sampling from a discrete uniform distribution of the $\varepsilon'_v$ set. Equation 4 guarantees that no community will ever be eliminated from the network since, when a community $c$ is compressed into a single node $v$, $c$ has the maximum possible density on the update rule of $v$ (i.e., 1.0) guaranteeing $c \in \varepsilon'_v$, and thus $c'(v) = c$.

With this highly scalable community detection algorithm, LuluChain dynamically detect and find the most efficient witness node cluster. This improves the network performance in terms of processing transactions asynchronously fast (even with the same bandwidth).

***LuluChain network communications.*** LuluChain relies on efficient network communications to ensure that ledger information spreads widely among nodes, and to ensure that transactions are validated by a wide enough sample of witness nodes. LuluChain structures communications as a gossip network for witness nodes to propa-

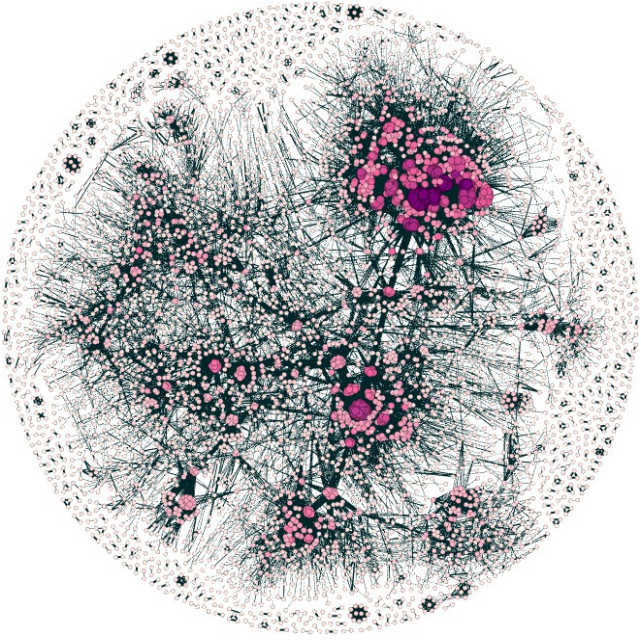

Fig. 4. Communities detected by asynchronous fluid detection algorithm. From here we choose the nodes in the largest community for the final set of witness nodes. Visualized with Fruchterman-Reingold graph drawing algorithm.

Adopting this network communities detection algorithm, LuluChain can dynamically find out the best set of witness nodes. Fluid Communities algorithm is based on the idea of introducing a number of fluids (i.e., communities) within a non-homogeneous environment, where fluids will expand and push each other influenced by the topology of the environment until a stable state is reached. Given a network graph G = (V, E) composed by a set of nodes V and set of links E, Fluid Communities initializes k fluid communities $\varepsilon = \{c_1, c_2, \dots c_k\}$, where $0 < k \leq |V|$. Each community $c \in \varepsilon$ is initialized in a different and random node $v \in V$. Each initialized community has an associated density d within the range (0,1]. The density of a community is the inverse of the number of nodes (1) composing said community.

$$d(c) = \frac{1}{|v \in c|} \tag{1}$$

Fluid Communities operates through supersteps. On each superstep, the algorithm iterates over all nodes of V in random order, updating community each node belongs to using an update rule. When the assignment of nodes to communities does not change on two consecutive supersteps, the algorithm has converged and ends. The update rule is formally defined in Equation (2) and (3).

$$\varepsilon'_v = argmax_{c \in \varepsilon} \sum_{w \in \{v, \Gamma(v)\}} d(c) \times \delta(c(w), c) \tag{2}$$

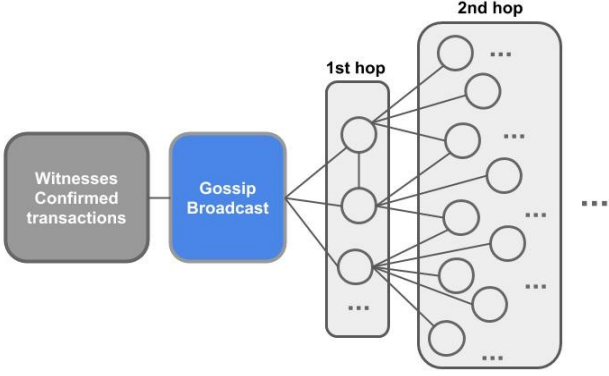

Fig. 6. Gossip protocol propagates the transaction acceptance/rejection messages across all the nodes in the network through out multiple hops.

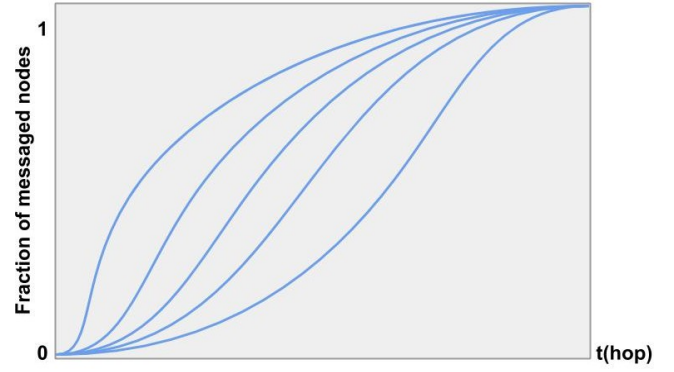

Fig. 7. Gossip protocol and characteristic time. When there are more starting witness nodes and higher fanout degree $f$, the message propagation is faster (from left lines to right lines).

gate transaction acceptance/rejection message to all other nodes. Gossip protocols (See. Fig. 6) have been shown to be highly efficient for broadcasting messages to all nodes in a network. Each node propagates messages to a subset of the full node population. A witness node set is chosen at random network mechanism to ensure fairness, security and redundancy so that attackers cannot predict and target the nodes that will carry a particular message. After confirming the proposed transaction, the chosen set of witness nodes are responsible for propagating acceptance/rejection messages across all nodes in the network. LuluChain gossip protocol model starts the message propagation from the witness nodes. Denote with $W(t)$ the number of nodes that already contain the acceptance/rejection message at time $t$ and with $U(t)$ the number of nodes that don't contain the message. At time $t = 0$, $N$ - number of witness nodes don't contain the message ($U(0) = N$ - number of witness nodes) and only the witness nodes contain the message ($W(0) =$ number of witness nodes). Let's assume that the fan-out degree of the message broadcasting is $< f >$ and that the likelihood that the message will be transmitted from a node that contains the message to a node who does not contain the message in a unit time is $\beta$. Within the homogeneous mixing hypothesis, the probability that the node that contains the message broadcasts the message to a node that does not contain the message is $U(t)/N$. Therefore the node that contains the message propagate the message to $< f > U(t)/N$ nodes that don't contain the message. Since $W(t)$ nodes are transmitting the message, each at rate $\beta$, the average number of nodes that newly contain the message $dW(t)$ during a timeframe(a hop) $dt$ is equated as shown below:

$$\frac{dW(t)}{dt} = \beta < f > \frac{U(t)W(t)}{N}. \qquad (5)$$

Let $u(t) = U(t)/N$ and $w(t) = W(t)/N$ to capture the fraction of the nodes that don't contain the message and that do at time $t$. Re-writing Eq.(5) gets:

$$\frac{du(t)}{dt} = \beta < f > u(t)w(t) = \beta < f > w(t)(1-w(t)), \quad (6)$$

$$\frac{du(t)}{u(t)} + \frac{du(t)}{(1 - u(t))} = \beta < f > dt, \qquad (7)$$

$$lnu(t) + ln(1 - u(t)) + C = \beta < f > t, \qquad (8)$$

With the initial condition $u_0 = u(t = 0)$, we get $C = u_0(1 - u_0)$, obtaining that the fraction of the nodes that contain the message increases in time as:

$$u(t) = \frac{u_0 e^{\beta < f > t}}{1 - u_0 + u_0 e^{\beta < f > t}} \qquad (9)$$

This shows that, at the beginning the fraction of the node that contain the message increases exponentially. Indeed, early on a node that contain the message propagate message more to the nodes that don't contain the message, hence the message can easily spread. The characteristic time required to reach an $1/e$ fraction of all the nodes that don't contain the message is:

$$\tau = \frac{1}{\beta < f >} \qquad (10)$$

Hence $\tau$ is the inverse of the speed with which the message spreads through all the nodes. Eq.(10) shows that increasing either the density of links $< f >$ or $\beta$ enhances the speed of the message and reduces the characteristic time (See Fig. 7). As the message spreads with hops, a node that contains the message broadcasts the message to fewer and fewer nodes that don't contain the message. Hence the growth of $u(t)$ slows for large $t$. The propagation ends when every node contains the message, i.e. when $u(t \to \infty) = 1$ and $u(t \to \infty) = 0$.

LuluChain incentivizes some nodes to serve as bridge nodes. The role of bridge nodes is to help nodes that are behind a firewall to communicate with the rest of the Lulu network. On initialization a node checks whether it is behind a stateful firewall that blocks incoming (push) connections. The node checks by sending an outbound test-request message to a bridge node. The bridge node tries to open an inbound connection to the node. If it succeeds, then the node is accessible and does not need a bridge. If the node fails to receive an inbound connection from the bridge node, then it opens outbound connections to one or more bridge nodes. The bridge nodes keep the connection open using periodic keep alive replies on the connection. When a bridge node needs to push a message to the node, it sends the message on the already-open connection - which a firewall normally treats as a response to a client request. Using this technique, nodes participate fully in the LuluChain network even behind a firewall.

***LuluChain infrastructure and architecture.*** The LuluChain implementation comprises two software modules: backend and frontend. The backend module is provided in multiple formats tailored to run on a variety of platforms, ranging from mobile devices through mainframe servers, and compiled for common operating systems including Windows, Macintosh, Linux, iOS, and Android. All backend formats implement the same functionality, but with differing platform-specific configurations and capacities. The backend module implements the LuluChain protocol, and manages platform-specific functions such as ledger storage and network communication. The backend is accessed via the LuluChain API, which is uniform across all platforms. The frontend module is also provided in multiple formats, for the same platforms and operating systems. The frontend implements optional end-user functionality, including GUI and personal wallet management. The frontend communicates with the backend via the Lulu API. An enterprise may develop its own LuluChain-based application software, using the API to communicate with instances of the LuluChain backend. A LuluChain network acts as a decentralized ledger platform-as-a-service (PaaS), available to applications for their custom usage. For as many platforms as possible, the LuluChain backend and frontend modules are provided in containerized packaging, which enhances security of the software through sandbox isolation, and bundles correct versions of system software to facilitate installation. LuluChain also provides a website which runs instances of the frontend, for customers who prefer to access their accounts via a web browser rather than an installed app.

***LuluChain security.*** LuluChain uses several methods to ensure that invalid transactions (such as double-spend or overdraw attempts) are rejected, and only valid transactions are accepted. LuluChain includes information in each transaction record which allows any computer node to detect an invalid proposed transaction. This method allows a witness node to compare the sender's ledger history to the proposed new spend, and to reject a double-spend or overdraw attempt without having to consult other nodes.

LuluChain validates the signatures of both the sender and receiver of every transaction. Because both parties sign, neither can repudiate the transaction. It is the sender's responsibility to ensure that enough funds are available in the sending account, and that no double-spend is attempted. If the LuluChain network detects a violation, the sender is on the hook for penalties, such as a slash fine. Because LuluChain is capable of detecting all violations and rejecting all invalid transactions, there is strong incentive for honesty. It is to the receiver's benefit to validate a transaction before attempting to use the received funds or to execute the reciprocal side of a trade (for example, delivering a purchased item). Thus the receiver initiates broadcast of the proposed transaction to witness nodes, for validation against their ledgers. Because LuluChain is not subject to total-ordering agreement protocols, it is extremely fast. For added security, a delay of several seconds is imposed on the transaction recipient before the funds can be spent. This gives the recipient extra safety in case a distant or delayed LuluChain node reports invalidity. LuluChain requires that witness nodes sign their transaction rejection/acceptance reports, and guarantee their honest performance via a stake. False or invalid witness reports are easily detected by other LuluChain nodes.

***Ledger alteration attacks.*** An attacker cannot rewrite a transaction record, because the record must be signed using the private keys of both the sender and receiver. The attacker cannot do that without both private keys. Even with both keys, an attacker cannot successfully broadcast a modified transaction, because the modification is detected and rejected by LuluChain witness nodes. An attacker could theoretically remove a record from a node that the attacker controls. But the attacker cannot remove the record from all nodes. And a single instance of the record signed by both parties, from any node, is sufficient evidence for the validity of the transaction. In case of two transaction records that include the same transaction identifiers, each signed by the same sender, but each with a different recipient, or a different amount, then the sender is at fault and is immediately detected and penalized. The first transaction may be preserved, and the second is rejected.

***Sybil attacks.*** LuluChain offers immunity to Sybil attacks - where an attacker creates multiple corrupt nodes that falsely report that an invalid transaction is valid. No matter how many such corrupt nodes are present, a single honest witness node in the network can verifiably report invalidity - in which case the transaction is rejected. Moreover, the false reports from the corrupt nodes are detected as invalid, which causes the corrupt nodes to be penalized and removed.

***Partition attacks.*** LuluChain also protects against network partition attacks. In a typical partition attack, an adversary gains control of network communications routing, and cuts one set of nodes off from the rest of the network. Then with the network partitioned into two sections that cannot communicate to each other, the adversary would spend an amount twice – sending it to a different recipient in each partition. The double-spend cannot be detected until the partition is healed. At that point it would be too late to prevent both recipients from having received and then further spent the amount. LuluChain uses three security strategies to detect and prevent partition attacks. First, LuluChain nodes monitor network connectivity, and if the number of connected nodes rapidly falls more than 50%, then nodes delay accepting new transactions until a sufficient number of the disconnected nodes reconnect, or until enough new nodes connect. The 50% threshold allows transactions to continue to complete in the larger partition. Second, LuluChain uses IP address geo-location to identify the global region of each node, and requires that connectivity to at least a minimum set of geographically dispersed regions be present. If a node loses connectivity with too many regions then it delays accepting new transactions until dispersed connectivity returns. Most partition attacks would occur in a localized region - for example, cutting off a country WAN or corporate LAN by taking control of its network routers. This method protects LuluChain nodes inside a localized partition by delaying transactions inside that partition until connectivity is restored. Third, in the unlikely event that a double-spend evades the first two partition security methods, LuluChain penalizes the double-spender by slashing funds remaining in the account.

***LuluChain fee and incentive structure.*** LuluChain transactions are intended to be as frictionless as possible. Transactions are nearly free to Lulu senders and receivers. A small percentage capped transaction fee, in the form of LuluChain debits, is charged to senders or receivers, to discourage wash transactions. The fee structure is designed to encourage both micro and macro transactions. LuluChain incentivizes nodes to perform community service for the benefit of the LuluChain network. The incentive is in the form of minted Lulu value. LuluChain allows two forms of community service: witnessing and bridging. The incentive is structured to discourage padding artificially the number of witnessed transactions. Every LuluChain node (except those disqualified for dishonesty) is eligible to receive community service incentive payments. This policy encourages nodes to stay online.

***LuluChain currency minting.*** LuluChain permits minting of LuluChain currency, under defined circumstances, as incentive to nodes for community service activities. Minting of allowed amounts is trivial to implement in LuluChain.

*Minting LuluChain currency for witness incentive.* A LuluChain node that wants to serve as a witness broadcasts a Witness Available message to other nodes. When a Witness Available message is received, the receiving node adds the new witness node to its list of available witness nodes. When a sender and receiver want to complete a transaction, a subset of the available witness nodes is selected for them. The selection is partly random, for probabilistic fairness, and partly determined by other criteria, including geographic dispersion, for security. The proposed receiver sends the dual-signed transaction record to the selected witness nodes. Each selected witness node then validates the transaction and, if it is valid, signs the transaction record, broadcasts it to the network, and stores it. Once per day, a LuluChain witness node can issue a special Witness Compensation transaction, payable to itself. The witness node is eligible to do so if it has served as witness for a minimum number of transactors during the day. The Witness Compensation transaction record attaches a copy of the witnessed, signed transaction records. It also includes the witness node's ID, and the standard daily witness-compensation fee amount. The witness node issues its Witness Compensation transaction by broadcasting the record to the LuluChain network (See Fig. 8). Other witness nodes that receive the Witness Compensation transaction record are programmed to check the validity of the witnessed transaction records that are attached, by confirming the signature pairs, and comparing against the ledgers for the sender and receiver. They also check the validity of the witness node's ID and the standard witness-compensation amount. If they detect an error, such as the witness node claiming to have witnessed an invalid transaction, then they sign and broadcast a rejection of the Witness Compensation, and penalize the witness node by adding it to a blacklist. If they detect no errors, then they sign and broadcast an acceptance message, as a witness. In this way, the Witness Compensation amount is effectively minted. LuluChain minting comprises just the witness node's issuance of the Witness Compensation transaction, followed by other nodes' signed acceptance, based on their validation checks. By allowing the validated Witness Compensation transaction to succeed, Lulu allows the witness node's balance to increase, without decreasing another node's balance. If a witness node fails to serve as a witness for any transactions for a given period of time, then the other nodes remove it from their list of available witness nodes. This scheme compensates witness nodes for their witness services by a standard, fixed amount per day, rather than in proportion to the transaction values. (A value-proportional scheme is not used, because that would incentivize wash transactions, to inflate witness compensation.) LuluChain's method of minting new currency, by accepting (under well-defined conditions) a credit that lacks a corresponding debit, eliminates the need for complex and burdensome coin-mining computations like Bitcoin's proof-of-work.

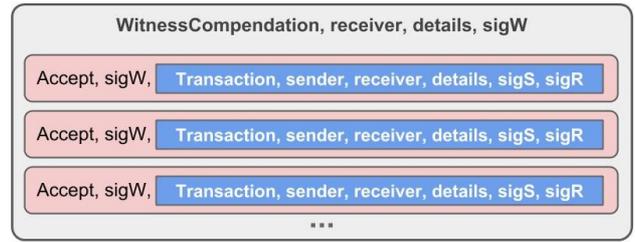

Fig. 8. Example of a Witness Compensation transaction record with witness records attached, each containing a transaction record where sigW, sigS and sigR mean signature of the witness, sender and receiver respectively.

***LuluChain privacy features.*** LuluChain provides the same privacy features as Bitcoin, plus two additional measures. Bitcoin privacy is essentially pseudonymity: a transaction sender or receiver is denoted by a unique hash code, which does not reveal the transactor's name or other identifying information. However, if a sender or receiver participates in multiple transactions, then those transactions can be linked together via their common hash code in the ledger. IP addresses and timestamps can also be used to link transactions together. The linkage of transactions may allow one or more transactors to be partially or fully identified.

*Dandelion++.* One of LuluChain's additional privacy measures is a variant of the Dandelion++ technique which has been proposed for Bitcoin. In Dandelion++, a transaction record is encrypted and then sent for several hops along a linear "stem" of nodes, alongside a random selection of other transaction records. At the end of the stem, a node broadcasts the record to a "fluff" of LuluChain wit-ness nodes. Because the record has been transported with other records originating from other nodes, an observer cannot determine the IP address from which the record originated. Dandelion++ thus obfuscates the IP address of a transaction, and prevents an observer from correlating IP addresses to identify transactors. Dandelion++ can also add small random delays in transmitting a record along the stem, making it difficult for an observer to correlate the timestamps of multiple transaction records originating from the same trade.

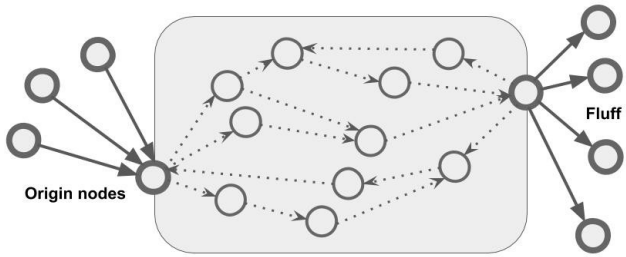

Fig. 9. sigS and sigR mean signature of the witness, sender and receiver respectively.

*Account spawning.* The second additional privacy measure that LuluChain provides is to require that all recipient accounts be new - i.e., that a user can receive funds only by creating a new account ID which starts with a zero balance. This means that every recipient account ID has only one sender. Compared to Bitcoin, in which a recipient account may receive transaction amounts from multiple senders, the "attack surface" for correlating LuluChain transactions is thus greatly reduced. An electronic wallet, part of LuluChain's frontend software module, serves as the user's agent to manage multiple account IDs. These additional privacy measures make it more difficult for an observer to infer or narrow the possible identity of a transactor. They thus enhance transactors' privacy.

*Further privacy measures.* LuluChain may implement further privacy measures, such as zero-knowledge proofs or CoinJoin, in the future.

## 4. COMMERCIAL APPLICATIONS

*Trade marketplaces and exchanges.*
With the fast throughput of LuluChain, it can serve as a trusted decentralized exchange - for stock trading, ad serving, idle computation leasing, and other marketplace applications. It can accommodate a high-volume order book without sacrificing speed.

*Enterprise processes.*
With its simplicity, LuluChain is particularly well suited for development and deployment of enterprise applications that use the LuluChain API. Less overhead and complexity is involved in building and operating distributed ledger applications than for competitors like Bitcoin and Ethereum. Such applications include banking account functions like deposits, withdrawals, transfers, clearing, settlement, cross-border wires, merchant incentives and promotions, and co-investment deals.

*Full range of devices.*
With its simplicity, LuluChain is also well suited for lightweight implementation that runs on a range of devices, from less-powerful mobile flip-phones, to personal wallet card devices, to smartphones, to tablet/laptop computers, to PCs - as well as implementation on full-size enterprise mainframe computers. Thus LuluChain can support a distributed ledger economy in economically emerging countries where low-end devices are common.

*Micropayments.*
With its low computing demands, LuluChain operates with extremely low transaction costs. Consequently it is well suited for frictionless microtransactions, with low percentage transaction fees, as well as larger transactions, with capped transaction fees.

*Rendering Microservices.*
CGI and video rendering microservices have great potential to fuel LuluChain's distributed cloud marketplace economy. The use case for CGI rendering is strong and preferable for CGI artists when compared with traditional cloud services. Artists will be able to rent idle computing power on LuluChain using LULU cash to quickly produce CGI or support video rendering. Conversely, idle machines can accept tasks from other users on the marketplace creating alternative, decentralized income streams.

*High-Level Data Analysis.*
Hospitals can benefit from LuluChain's high-level data analysis and the distributed system's electronic medical record (EMR) storage capacity. Secure and immutable storage coupled with supercomputing power allow all EMRs updated in real-time and automatically analyzed. Medical institutions can use perpetually updated and analyzed data pools to improve medical intervention strategies.

## 5. CONCLUSION

This paper describes a high-throughput blockchain secured via ledger separation that serves as an enterprise-grade developer sandbox and is backed by a powerful supercomputing engine. This provides a robust functional basis for a digital cash-based distributed cloud ecosystem. LuluChain has a novel, leaderless, and highly scalable PKI-signed based random consensus driven network. LuluChain has the necessary infrastructural properties to scale high-speed, easy-to-integrate decentralized applications and accommodate collective computational processing for all major industries. The integrity of Witness Credits as a fuel to run a cohesive network where both the witness and governing body earn LULU cash at the same time enables LuluChain to have a secure, scalable, and privacy centric network that will effectively shift the cloud computing paradigm. LuluChain's emphasis on alleviating compliance stressors creates a plug-and-play utopia for enterprise-grade developers, ultimately making it the most attractive development platform with a truly decentralized cloud-based economy.